\documentclass{acm_proc_article-sp}
\usepackage{pgf}
\usepackage{enumerate}
\usepackage{graphicx, psfrag}
\usepackage[labelsep=period,font=bf]{caption}
\usepackage{subcaption}
\usepackage[font=bf]{subfig}
\usepackage{amsmath}
\usepackage{amssymb, dsfont, bm, mathrsfs} 
\newtheorem{thm}{Theorem}[section]

\newtheorem{prop}[thm]{Proposition}
\newtheorem{defn}[thm]{Definition}

\newcommand{\set}[1]{\left\{#1\right\}}

\newcommand{\trr}{\triangleright}
\newcommand{\rrt}{\triangleleft\,}

\newcommand{\ass}{\stackrel{\textup{\tiny def}}{=}}

\renewcommand{\theenumi}{\roman{enumi}}

\newcounter{saveenumerate} 
\makeatletter
\newcommand{\enumeratext}[1]{%
\setcounter{saveenumerate}{\value{enum\romannumeral\the\@enumdepth}}
\end{enumerate}
#1
\begin{enumerate}
\setcounter{enum\romannumeral\the\@enumdepth}{\value{saveenumerate}}%
}

\makeatletter
  \let\@copyrightspace\relax
  \makeatother

\begin{document}

\title{Low-Dimensional Topology of Information Fusion}

\numberofauthors{2} 
%
\author{
\alignauthor
Daniel D. Moskovich \\
\affaddr{School of Physical and Mathematical Sciences} \\
\affaddr{Nanyang Technological University} \\
\affaddr{Singapore} \\
\email{dmoskovich@gmail.com}
\alignauthor
Avishy Y. Carmi \\
\affaddr{Faculty of Engineering Sciences} \\
\affaddr{Ben-Gurion University of the Negev} \\
\affaddr{Israel} \\
\email{avcarmi@bgu.ac.il}
}



\date{}



\maketitle
\begin{abstract}
We provide an axiomatic characterization of information fusion, on the basis of which we define an \emph{information fusion network}. Our construction is reminiscent of tangle diagrams in low dimensional topology. Information fusion networks come equipped with a natural notion of \emph{equivalence}. Equivalent networks `contain the same information', but differ locally. When fusing streams of information, an information fusion network may adaptively optimize itself inside its equivalence class. This provides a fault tolerance mechanism for such networks.
\end{abstract}

\keywords{information fusion, information entropy, adaptive sensor networks, estimation theory, covariance intersection, fault tolerance, Reidemeister moves, tangle diagrams}

\section{Introduction}\label{S:Introduction}

\emph{Information fusion} is an umbrella term for concepts and methodologies whose primary goal is to integrate heterogeneous pieces of information from diverse sources. An information fusion algorithm is a vital ingredient of virtually any autonomous system~\cite{fuse3, fuse9, fuse1,Murphy:00}. Some other applications are sensor networks~\cite{fuse4, fuse4a}, biometrics~\cite{fuse5}, and intelligent decision support systems~\cite{fuse8,fuse6}.
Here we consider pieces of information whose correlations are unknown and some of which may originate from unreliable sources. `Pieces of information' may refer to estimates of unknown parameters or state variables, or to other related statistical measures such as (unnormalized) probability density functions, Fisher information and Shannon entropy.

We axiomatize information fusion (Definition~\ref{D:Quandloid}), and we define an \emph{information fusion network} (Definition~\ref{D:UpdateNetwork}). A similar concept of a \emph{tangle machine} was defined in \cite{Carmi:14a}. Information fusion networks admit a natural notion of equivalence (\ref{S:Similarity}). When one or more streams of information becomes faulty (\textit{e.g.} biased or inconsistent), the faults will propagate differently in equivalent networks. At every moment, an information fusion network may be adaptively configured to its `least faulty' equivalent configuration. Examples are given in Sections~\ref{S:example} and~\ref{S:Resilience}.

 Our notion of equivalence parallels the notion of \emph{ambient isotopy} in knot theory. As such, it represents a link between information fusion and low dimensional topology. As discussed in \cite{Carmi:14b}, low dimensional topology provides insight into the characteristic quantities or \emph{invariants} of information fusion networks, such as how much information is required to uniquely specify the network (its \emph{capacity}) and how many `independent sub-networks' the network contains (its \emph{complexity}).

\section{An illustrative example}\label{S:example}


\subsection{The example}\label{SS:example}
This example illustrates two different \emph{information fusion networks} for fusing estimates. These networks are \emph{equivalent} in the sense that any information in one network can uniquely be recovered from the information in the other, but they differ in the consistency of intermediate fused estimates. Thus, one network might be `better' than the other.

Consider three estimators $\hat{X}_0$, $\hat{X}_1$ and $\hat{X}_2$ for the same random variable $X$. The correlation between these estimators is unknown. We are also provided with estimators $C_0$, $C_1$, and $C_2$ for the error covariances:
\begin{multline}
\mathrm{cov}\left[ X-\hat{X}_i \right] \ass
E\left[(X-\hat{X}_i)(X-\hat{X}_i)^T \right] \\ - E\left[X-\hat{X}_i\right] E\left[X-\hat{X}_i\right]^T \quad \text{$i=0,1,2$,}
\end{multline}
\noindent where $E\left[ \,\bullet\, \right]$ denotes the expectation taken with respect to the underlying joint probability distribution.

A \emph{consistent estimator} $(\hat{X}_i, \; C_i)$, also called a \emph{conservative estimator}, is one which is not \emph{too} optimistic about its belief of what the value of $X$ is~\cite{CI1}:
\begin{equation}
C_i  \ge \mathrm{cov}\left[ X-\hat{X}_i \right]
\end{equation}
\noindent \textit{i.e.} the matrix difference $C_i  - \mathrm{cov}\left[ X-\hat{X}_i \right]$ is positive semi-definite. We would like all estimators to be consistent because inconsistent estimators may diverge and cause errors.

\emph{Covariance intersection} (CI) provides a method for fusing \textbf{a pair} of consistent estimates whose correlations are unspecified~\cite{CI1,CI2}. The working principle of CI is that if $\hat{X}$ and $\hat{X}^\prime$ are consistent estimators of $X$, then so is their convex combination (the proof is provided in \cite{CI1}):
\begin{subequations}
\label{eq:ci}
\begin{equation}
\hat{X}_{a} = (1-s) C_{a} C_0^{-1} \hat{X}_0 + s C_a C_1^{-1} \hat{X}_1,
\end{equation}
\begin{equation}
C_a^{-1} = (1-s) C_0^{-1} + s C_1^{-1},
\end{equation}
\end{subequations}
where $s \in [0,1)$ is a weight parameter. Different choices of the parameter $s$ can be used to optimize the update with respect to different performance criteria.

Our goal is to fuse \textbf{the triple} $(\hat{X}_0, C_0)$, $(\hat{X}_1, C_1)$, and $(\hat{X}_2, C_2)$ to obtain a single consistent estimator for $X$. Two fusion schemes are:

 \begin{enumerate}
 \item First fuse $(\hat{X}_0, C_0)$ with $(\hat{X}_1, C_1)$ with a parameter $s$, then fuse the resulting estimator with $(\hat{X}_2, C_2)$ with a parameter $t$.
 \item First fuse both $(\hat{X}_0, C_0)$ and $(\hat{X}_1, C_1)$ with $(\hat{X}_2, C_2)$ with the parameter $t$. Then fuse the resulting fused estimators with the parameter $s$.
 \end{enumerate}

Later, we will represent these two fusion schemes by Figure~\ref{F:FusionNets}. Pairs are fused using CI using the same weights $s$ and $t$. A short computation confirms that both of the above fusion schemes for a consistent triple $(\hat{X}_0, C_0)$, $(\hat{X}_1, C_1)$, and $(\hat{X}_2, C_2)$ result in the same consistent estimator for $X$. 

But what if $(\hat{X}_0, C_0)$ and $(\hat{X}_2, C_2)$ were consistent, but $(\hat{X}_1, C_1)$ was inconsistent? Then both fusion schemes \textbf{ultimately} lead to the same estimate of $X$, which is consistent for an appropriate choice of $s$ and of $t$. But \textbf{at an intermediate stage}, the first fusion scheme involves the inconsistent estimate ``$(\hat{X}_0, C_0)$ fused with $(\hat{X}_1, C_1)$ with parameter $s$'', whereas all intermediate fused estimates in the second scheme are consistent. In this case the second fusion scheme is \emph{better} than the first. Conversely, if $(\hat{X}_1, C_1)$ is highly consistent compared to $(\hat{X}_0, C_0)$ and $(\hat{X}_2, C_2)$, then the first fusion scheme would be better than the second.

\subsection{The information fusion networks of our example}\label{SS:NetworkExample}

\begin{figure*}
\renewcommand{\thesubfigure}{\Alph{subfigure}}
\centering
\begin{subfigure}{.48\textwidth}
  \centering
  \psfrag{0}[c]{\small $0$}
\psfrag{1}[c]{\small $1$}
\psfrag{2}[c]{\small $2$}
\psfrag{a}[c]{\small $a$}
\psfrag{b}[c]{\small $b$}
\psfrag{c}[c]{\small $c$}
\psfrag{d}[c]{\small }
\psfrag{s}[c]{\small $\trr_s$}
\psfrag{t}[c]{\small $\trr_t$}
\includegraphics[width=0.65\textwidth]{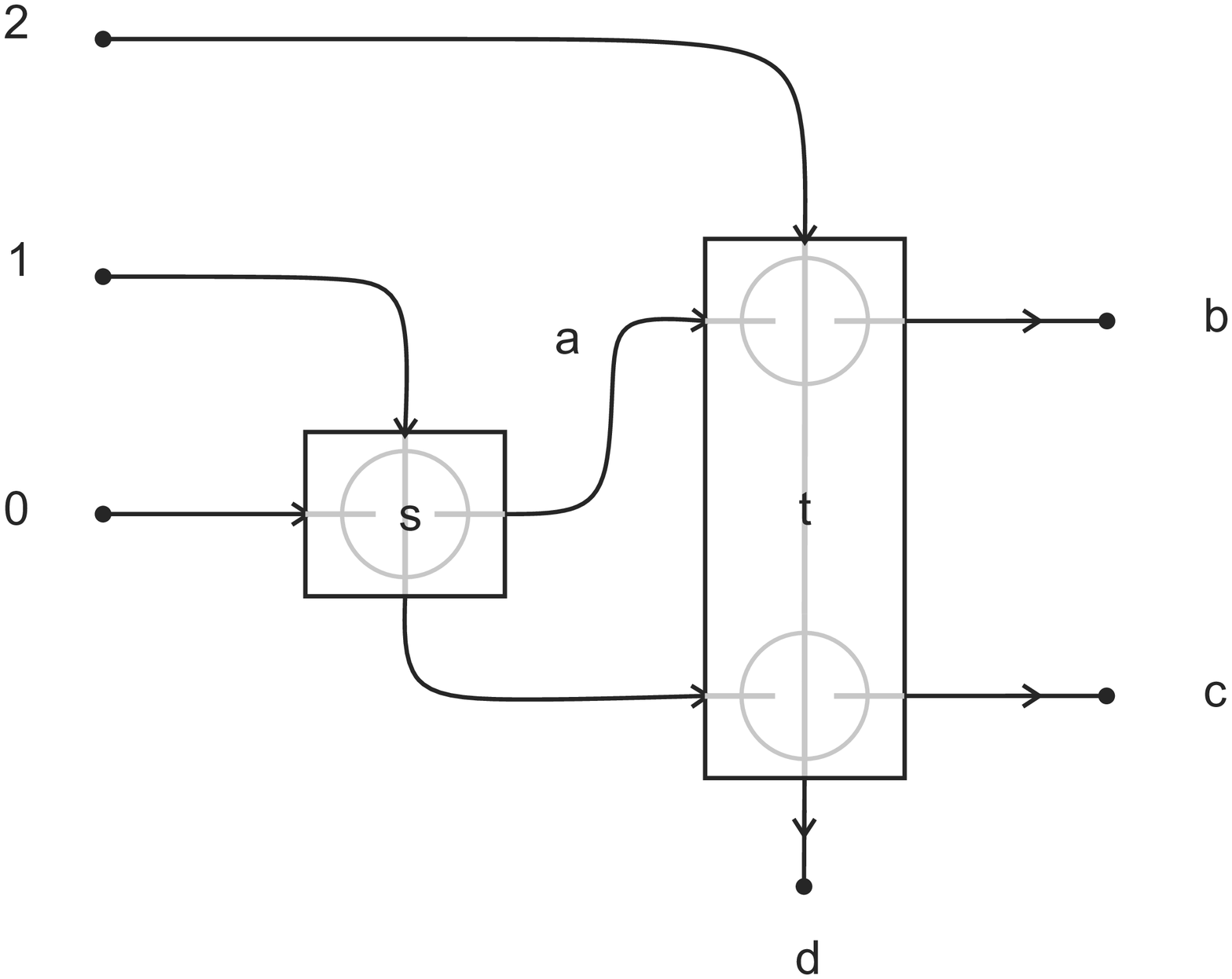}
\caption{\label{F:netsub1}}
\end{subfigure}%
\begin{subfigure}{.48\textwidth}
  \centering
  \psfrag{0}[c]{\small $0$}
\psfrag{1}[c]{\small $1$}
\psfrag{2}[c]{\small $2$}
\psfrag{a}[c]{\small $\bar{a}$}
\psfrag{b}[c]{\small $\bar{b}$}
\psfrag{c}[c]{\small $c$}
\psfrag{d}[c]{\small }
\psfrag{s}[c]{\small $\trr_s$}
\psfrag{t}[c]{\small $\trr_t$}
\includegraphics[width=0.65\textwidth]{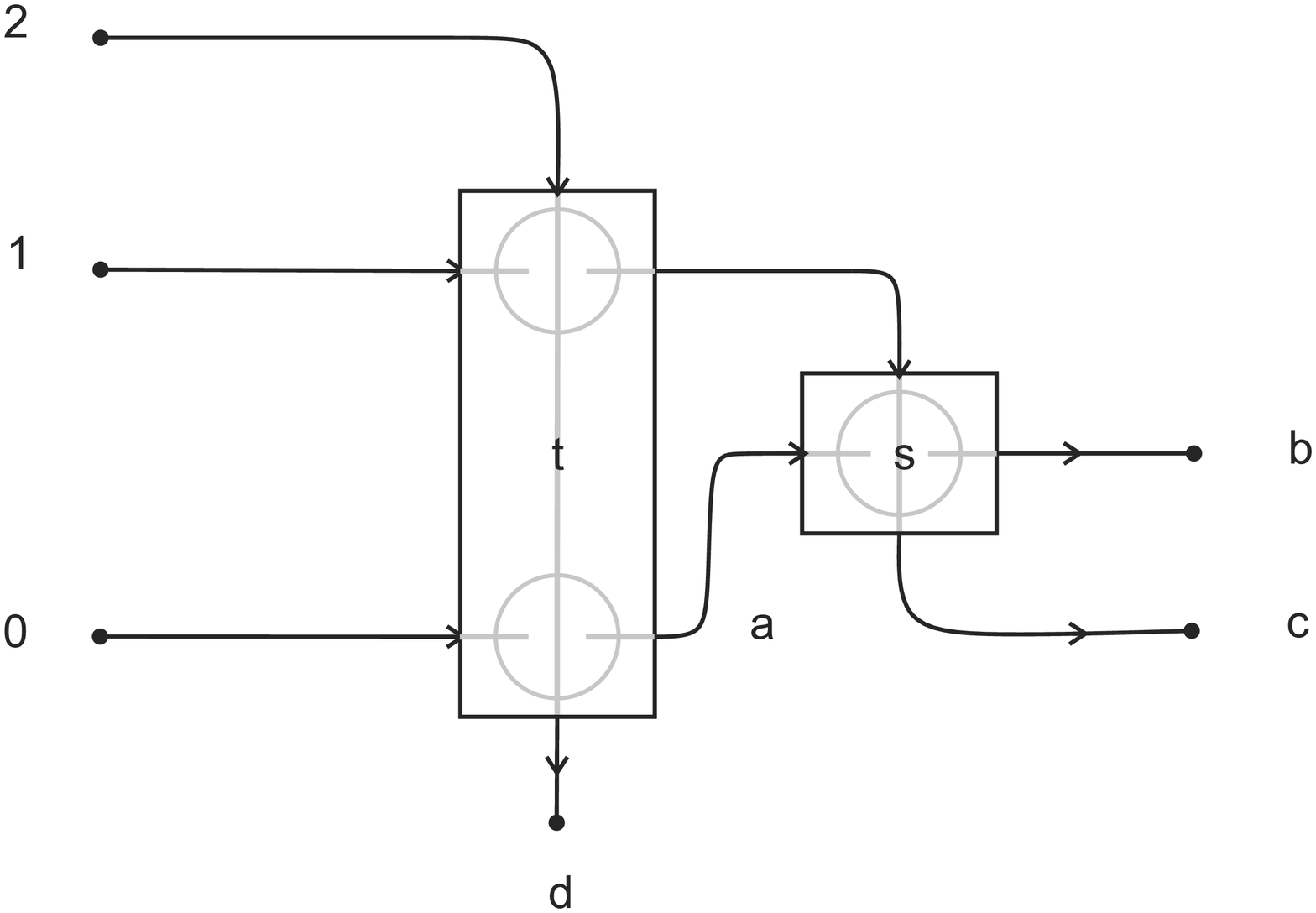}
\caption{\label{F:netsub2}}
\end{subfigure}
\caption{\label{F:FusionNets} \small Subfigures \ref{F:netsub1} and \ref{F:netsub2} represent different schemes to fuse estimates $(\hat{X}_0, C_0)$, $(\hat{X}_1, C_1)$, and $(\hat{X}_2, C_2)$ to a consistent estimate for $X$.} 
\end{figure*}

This section constitutes a preliminary discussion of the concept of an information fusion network in the context of our example, in anticipation of its definition in Section~\ref{S:IFN}. An \emph{information fusion network} is made up of labeled directed edges whose incident nodes may either be fixed points in the plane called \emph{endpoints} or boxes called \emph{interactions}. 
Each interaction involves one distinguished edge called the \emph{agent} which goes `all the way though' vertically and `emerges on the other side' (\textit{i.e.} we identify the edge incident to the top to the bottom of the interaction). Other edges are called \emph{patients}, and come in pairs--- there is one patient coming into the interaction for each patient going out of the interaction. 

Figure~\ref{F:FusionNets} contains two subfigures, representing the two information fusion schemes in Section~\ref{SS:example}. Each have three \emph{input edges}, labeled $0$, $1$, $2$, denoting estimates $(\hat{X}_0, C_0)$, $(\hat{X}_1, C_1)$, and $(\hat{X}_2, C_2)$ correspondingly, and a distinguished output edge labeled $b$ in \ref{F:netsub1} and $\bar{b}$ in \ref{F:netsub2}. Their interactions are labeled by operations $\trr_s$ for some $s\in \mathds{R}$, which represents an application of the fusion rule \eqref{eq:ci} with weight $s$. Explicitly, agents are weighted $s$ and patients are weighted $1-s$.

A \emph{fusion process} is a sequence of estimates stored in edges going from an initial to a terminal edge. Subfigure~\ref{F:netsub1} contains three fusion processes, labeled $\{0,a,b\}$, $\{1, c\}$, and $\{2\}$.  We write $0 \trr_s 1$ for the fusion of the estimate $(\hat{X}_0, C_0)$ with $(\hat{X}_1,C_1)$ according to the fusion rule \eqref{eq:ci} with a weighting parameter $s$. The resulting estimate $(\hat{X}_a, C_a)$ is then passed on to the second interaction where it fuses with $(\hat{X}_2, C_2)$ with parameter $t$, and resulting estimate $(0\trr _s 1)\trr_t 2$ is stored in $b$.


The network of Subfigure~\ref{F:netsub2} differs only in its first process, which reads:
\begin{equation}
0 \rightarrow \underbrace{0 \trr_t 2}_{\bar{a}} \rightarrow \underbrace{(0 \trr_t 2) \trr_s (1 \trr_t 2)}_{\bar{b}}
\end{equation}



Above we described a pair of equivalent information fusion networks which have different local performance. Alternating between these two networks to match a time evolution of initial estimator consistencies adaptively optimizes the consistency of the intermediate estimator at each point in time. In future sections we consider the general case.


\section{Algebra of information fusion}\label{SS:Quandloid}

Consider a sensor network~\cite{fuse4, fuse4a}. In distributed information fusion architectures, nodes behave as intelligent proxies fusing raw measurements streaming from their sensors with information received from neighboring nodes~\cite{fuse4a}. Fusion may be carried out within a node using a statistical filtering algorithm, \textit{e.g.} the Kalman filter. Such algorithms normally use cross dependencies between the incoming pieces of information (\textit{i.e.} raw measurements and estimates from other nodes). However, large scale and complex networks generally inhibit calculations of the required statistical interdependencies among nodes~\cite{CI1}. Covariance intersection (CI) does not require knowledge of correlations, hence it is suitable for fusion in large scale settings~\cite{CI1,CI2}. Definition~\ref{D:Quandloid} abstracts the key properties of CI to axiomatize information fusion. Chief among these is self-distributivity or \emph{no double counting} which guarantees equality of outputs of the two networks of Section~\ref{S:example}.


\subsection{Algebraic axiomatization}

In this section we formulate an algebra structure of `information' subject to a binary `update' operation. We name such a structure a \emph{quandloid}, a portmanteau of ``quandle'' and ``groupoid''. 



\begin{defn}[Quandloid]\label{D:Quandloid}
A \emph{quandloid} is a set $Q$, whose elements, called \emph{colours}, represent pieces of information, together with a set $B$ of partially-defined binary operations representing `updates', satisfying the following properties:
\begin{itemize}
\item \emph{Coherence of information}. A piece of information may update itself, and this neither generates new information nor loses information. Symbolically:
    \begin{equation}
    a\trr a=a \qquad \text{$\forall a\in Q$, \quad $\forall\trr \in B$.}
    \end{equation}
\item \emph{Causal invertibility}. All update operations are left invertible. Thus, any input can uniquely be recovered from its corresponding output together with the agent. Symbolically, for every $\trr\in B$ there exists an `left-inverse operation' $\rrt\in B$ such that:
    \begin{equation}
    (a\trr b)\rrt b =a \qquad \text{$\forall a,b\in Q$, \quad $\forall \trr\in B$.}
    \end{equation}
    More precisely, if $a\trr b$ exists then $(a\trr b)\rrt b$ exists and equals $a$.
\item \emph{No double counting}. Updating $a\trr_2 c$ by $b\trr_2 c$ with $\trr_1$ gives the same result as the updated piece of information $a\trr_1 b$ by $c$ with $\trr_2$. Thus $c$ counts towards the final result only once, and there is no redundancy. Symbolically:
    \begin{multline}
    (a\trr_1 b)\trr_2 c= (a\trr_2 c)\trr_1(b\trr_2 c) \qquad \text{$\forall a,b,c\in Q$},  \\
\text{$\forall\trr_1,\trr_2\in B$.}
    \end{multline}
    In particular, the left-hand side exists if and only if the right-hand side exists.
\item \emph{Identity}. The set $B$ includes an `identity element' $\trr_e$ such that $a\trr_e b=a$ for all $a,b\in Q$.
\end{itemize}
By abuse of notation, we often write $Q$ as a shorthand for $(Q,B)$.
\end{defn}

Similar self-distributive structures have been studied in knot theory (\textit{e.g.} \cite{Carter:09}) and in the theory of computation (\textit{e.g.} \cite{Roscoe:90}). Specifically, if we were to require that all binary operations in $B$ be fully defined--- \textit{i.e.} that $a\trr b$ exists for all $a,b\in Q$ and for all $\trr\in B$ then our notion would recover the notion of a multi-quandle \cite{Przytycki:11}. 

\subsection{Linear and log-linear quandloids}

From Equation~\ref{eq:ci}, we see that an example of a quandloid is the set $Q$ of all pairs $(\hat{X},C)$ of an estimator $\hat{X}$ for a random variable $X$ with an estimator $C$ for its error covariance matrix, whose update operations are $\trr_s$ with $s\in [0,1)$ and their formal inverses $\rrt_s$ (where defined). Generalizing this example, we make a definition.

\begin{defn}[Linear quandloid]\label{D:SLQ}
Equipping a real vector space $\bar{Q}$ with operations $\bar{\trr}_s$ such that:
\[
a \, \bar{\trr}_s \, b \ass (1-s) a + s b, \qquad \text{$a,b \in \bar{Q}$,}
\]
with $s \in S\subset\mathds{R}$ where $1\notin S$ (by causal invertibility) but $0\in S$ (by identity) defines a quandloid called a \emph{linear quandloid}.
\end{defn}

A second class of examples of quandloids is given below:

\begin{defn}[Log-linear quandloid]\label{D:LSLQ}
Let $Q$ be a space of measures defined over the $\sigma$-algebra of some set $R$, whose elements we think of as unnormalized probability density functions. Then equipping $Q$ with the operations
\[
p(x) \trr_s q(x) \ass p(x)^{1-s} q(x)^s, \qquad \text{$p(x),q(x) \in Q$,}
\]
again with $s \in S\subset\mathds{R}$ where $1\notin S$ (by causal invertibility) but $0\in S$ (by identity) defines a quandloid called a \emph{log-linear quandloid}.
\end{defn}

Our next quandloids will be derived from log-linear quandloids using \emph{homomorphisms}.

\begin{defn}
A \emph{homomorphism} of quandloids is a function $\bm{h}\ass (h_E,h_V)\colon\, (Q,B)\to (\bar{Q},\bar{B})$, such that
\[
h_E(a \trr b) = h_E(a) \, h_V(\trr) \, h_E(b)\qquad \text{$\forall a,b\in Q$, \quad $\forall \trr\in B$.}
\]
\end{defn}


\subsection{Covariance intersection}\label{SS:CI}

The covariance intersection update rules are obtained via a homomorphism from a particular log-linear quandloid. Explicitly:

\begin{prop}
\label{thm:ci}
Let
\begingroup\makeatletter\def\f@size{7}\check@mathfonts
\def\maketag@@@#1{\hbox{\m@th\large\normalfont#1}}%
\begin{equation}
Q \ass \left\{ \left.\exp\left(-\frac{1}{2} (X - \hat{X})^T C_X^{-1} (X - \hat{X}) \right) \, \right| \, \hat{X} \in \mathds{R}^n\!, \; C_X \in \mathds{R}^{n \times n} \right \}
\end{equation}
\endgroup
\noindent be the space of unnormalized Gaussian probability density functions over $\mathds{R}^n$. The underlying space of parameters $\bar{Q} = (\mathds{R}^n, \mathds{R}^{n \times n})$ together with the operations $\bar{\trr}_s\colon\, \bar{Q} \times \bar{Q} \to \bar{Q}$
\begin{subequations}
\small
\begin{equation}
\label{E:21}
\hat{Z} = \left(C_Z C_X^{-1} \hat{X}\right) \bar{\trr}_s  \left(C_Z C_Y^{-1} \hat{Y}\right) = (1-s) C_Z C_X^{-1} \hat{X} + s C_Z C_Y^{-1} \hat{Y}
\end{equation}
\begin{equation}
\label{E:22}
C_Z^{-1} = C_X^{-1} \bar{\trr}_s  C_Y^{-1} = (1-s) C_X^{-1} + s C_Y^{-1}
\end{equation}
\end{subequations}
for every $(\hat{X}, C_X), \, (\hat{Y}, C_Y) \in \bar{Q}$, with $s\in [0,1)$, form the linear quandloid of estimators and CI, considered at the beginning of this note.
\end{prop}

The proof, which is straightforward, is deferred to the Appendix.

\subsection{Fisher information}\label{SS:Fisher}

Let $I_{\theta}$ be the \emph{observed} Fisher information matrix associated with the unnormalized joint probability density function:
\[
(p\trr_s q)(X, \theta) \ass p(X, \theta) \trr_s q(X, \theta)
\]
where $\theta$ is a random parameter vector whose prior is $g(\theta)$, and where $p(X, \theta)=p(X \mid \theta)g(\theta)$, $q(X, \theta)=q(X \mid \theta)g(\theta)$. Then $I_{\theta}$ satisfies:
\begin{multline}
\label{eq:ghom}
\underbrace{ - \frac{\partial^2 \log (p\trr_s q)(X, \theta)}{\partial \theta_i \partial \theta_j}}_{I_{\theta}(p \, \trr_s \, q)} = \\
\underbrace{\left( - \frac{\partial^2 \log p(X, \theta)}{\partial \theta_i \partial \theta_j}\right)}_{I_{\theta}(p)} \, \bar{\trr}_s \, \underbrace{\left( - \frac{\partial^2 \log q(X, \theta)}{\partial \theta_i \partial \theta_j}\right)}_{I_{\theta}(q)}
\end{multline}
where:
\[
a \, \bar{\trr}_s b \ass (1-s)a +sb, \qquad \text{$a,b \in Q$,}
\]
is a binary operation for a linear quandloid. Note that, in the case of a random parameter vector, the normalizing factor is independent of the parameters, and hence we may use the unnormalized $p\trr_s q$ in the expression for the observed Fisher information matrix.


Note that \emph{expected} Fisher information matrices do not form a quandloid in general. Exceptional cases, when expected Fisher information matrices do form a quandloid, are either when $I_{\theta} = E\{I_{\theta}\}$ or when the underlying expectation is taken exclusively with respect to the prior $g(\theta)$. We give examples of both of these cases.

Assume that $p(X, \theta)= p(X \mid \theta)g(\theta)$ and $q(X, \theta) = q(X \mid \theta)g(\theta)$ are colours of a log-linear Gaussian quandloid $Q$, where $\theta$ is a mode whose prior $g(\theta)$ is also Gaussian. Then $h_E(\cdot) \ass I_{\theta}(\cdot)$ is a homomorphism and $(E\{\mathcal{I}_{\theta}\}, \bar{\trr}_s)$ is a linear quandloid:
\begin{multline*}
E\left\{I_{\theta}(p \trr_s q)\right\} = (C_p^{-1} + C_g^{-1}) \, \bar{\trr}_s (C_q^{-1} + C_g^{-1}) \\ = E\left\{I_{\theta}(p)\right\} \bar{\trr}_s E\left\{I_{\theta}(q)\right\}
\end{multline*}
where $C_p$,$C_q$ and $C_g$ denote the covariances of $p$, $q$ and $g$, and $s\in[0,1)$.

Here is another example. Consider a log-linear quandloid whose colours are of the form $p(X, \theta)= p(X \mid \lambda_p(\theta))g(\theta)$ and $q(X, \theta)= q(X \mid \lambda_q(\theta))g(\theta)$ where the underlying conditionals are exponential densities whose $\theta$-dependent rate parameters are $\lambda_p$ and $\lambda_q$. As before $g(\theta)$ denotes the prior of $\theta$. Additionally, we assume that the second derivative of any rate parameter vanishes $d^2\lambda(\theta)/ d\theta^2 = 0$. In such a case, indeed:
\[
E\left\{I_{\theta}(p \trr_s q)\right\} = E\left\{I_{\theta}(p)\right\} \, \bar{\trr}_s E\left\{I_{\theta}(q)\right\}
\]
and thus $(E\{\mathcal{I}_{\theta}\}, \set{\bar{\trr}_s})$ a linear quandloid.


\subsection{Shannon information}\label{SS:Shannon}

This example was discussed in \cite{Carmi:14a}.

A binary information source is described by a Bernoulli random variable $X$. The entropy of $X$, denoted $H(X)$, has an operational meaning given by Shannon's source coding theorem. Very long independent identically distributed sequences generated by such a source
fall into two categories: they are either \emph{typical} or not. The probability of a typical sequence stabilizes around the value $2^{-N H(X)}$ which implies that not more than $N H(X)$ bits are required to encode any typical message with `negligible' loss of information.

Let $Q$ be a log-linear quandloid whose colours are uniform probability densities of typical sequences associated with infinitely many information sources
\[
Q \ass \left \{ 2^{-N H(X)} \mid X \text{ is an information source} \right \}
\]
The set of entropies $\{H(X)\}$ with operations $\bar{\trr}_s$ constitute a linear quandloid:
\begin{multline*}
H(Z) = H(X) \, \bar{\trr}_s H(Y) \\ = (1-s) H(X) + s H(Y), \qquad H(X), H(Y) \in \mathcal{H}
\end{multline*}
which follows from the homomorphism $h_E(\cdot) \ass - \frac{1}{N} \log (\cdot)$.


An update in this quandloid may describe a source $Z$ which behaves like $X$ in $(1-s)$ fraction of times and like $Y$ in the rest. For example, generating a message of $N$ symbols from $Z$, one expects that approximately $(1-s)N$ of them will make up a typical sequence from $X$ whereas the remaining $sN$ symbols will look as though they where generated by $Y$. In other words, the source $Z$ concatenates parts of the message from $X$ with parts of the message from $Y$.  



\section{What is an information fusion network?}\label{S:IFN}

%

We assemble information updates into a network. The philosophical position underlying its definition is that non-associative self-distributive algebraic structures (\emph{e.g.} quandloids) should not label graphs, but rather they should label low-dimensional `tangled' topological objects. 

\begin{defn}[Information fusion network]\label{D:UpdateNetwork}
An \emph{information fusion network} is:
\begin{itemize}
\item A directed graph $G$ whose vertices (drawn as boxes \textit{e.g.} in Figure~\ref{F:FusionNets}) have either degree $1$ or have positive even degree. Vertices of even degree are called \emph{interactions}, and the in-degree of each interaction equals its out-degree. Vertices of degree $1$ are called \emph{endpoints}. An endpoint is \emph{initial} if it is a source, and is \emph{terminal} if it is a sink.

\item For each interaction $v$, whose degree we denote $2m+2$, a partition of edges incident to $v$ into pairs: \[\set{(f^{\text{in}},f^{\text{out}})}\cup\set{(e_1^{\text{in}},e_1^{\text{out}}),(e_2^{\text{in}},e_2^{\text{out}}),\ldots (e_m^{\text{in}},e_m^{\text{out}})}.\]
    We consider each of the pairs $(f^{\text{in}},f^{\text{out}})$ as though it were a single edge, and refer to that pair as the \emph{agent} of $v$, and we call each $e_i^{\text{in}}$ an \emph{input} of $v$, and its corresponding edge $e_i^{\text{out}}$ an \emph{output} of $v$. All edges with superscript `in' (\textit{i.e.} $f^{\text{in}}$ and $e_1^{\text{in}},e_1^{\text{in}},\ldots,e_m^{\text{in}}$) are directed towards $v$, and all edges with superscript `out' are directed away from $v$.

\item A \emph{colouring function} $\bm{\rho}\ass (\rho_{V},\rho_E)$ where we define $\rho_V\colon\, V(G)_{\deg\geq 2}\to B$ and $\rho_E\colon\, E(G)\to Q$, where $(Q,B)$ is a quandloid. Function $\bm{\rho}$ must satisfy:
    \begin{enumerate}
    \item If $(f^{\text{in}},f^{\text{out}})$ is the agent of $v$, then $\rho_E(f^{\text{in}})= \rho_E (f^{\text{out}})$.
    \item For each input-output pair $(e_i^{\text{in}},e_i^{\text{out}})$ we have either $\rho_E(e_i^{\text{out}})= \rho_E(e_i^{\text{in}})\trr \rho_E(f)$ or else $\rho_E(e_i^{\text{in}})= \rho_E(e_i^{\text{out}})\trr \rho_E(f)$ where $\trr$ denotes $\rho_V(v)$, and $\rho_E(f)=\rho_E(f^{\text{in}})=\rho_E(f^{\text{out}})$ is the colour of the agent of $v$.
    \end{enumerate}
    Each transition $\rho_E(e_i^{\text{in}})\mapsto \rho_E(e_i^{\text{out}})$ is called an \emph{update} if $\rho_E(e_i^{\text{out}})= \rho_E(e_i^{\text{in}})\trr \rho_E(f)$ and is called a \emph{discount} if $\rho_E(e_i^{\text{in}})= \rho_E(e_i^{\text{out}})\trr \rho_E(f)$.
\end{itemize}
\end{defn}

Our diagrammatic convention is to update from right to left of the agent edge(s), so that if the agent edge is drawn pointing from top to bottom then the colour of an edge to the left of a box is updated by $\trr$ to become the colour of the corresponding edge the right of the box, and if the agent edge is drawn pointing from bottom to top then the opposite.

Note that the number of initial endpoints of an information fusion network equals its number of terminal endpoints. This relates information fusion to reversible computation, thanks to causal invertibility \cite{Land:61}.

If $(Q,B)$ were a multi-quandle, then the definition of an information fusion network is equivalent to the definition of a tangle machine in \cite{Carmi:14a}.

\section{Network equivalence}\label{S:Similarity}

\begin{figure*}
\renewcommand{\thesubfigure}{R\arabic{subfigure}}
\begin{subfigure}{.4\textwidth}
  \centering
  \psfrag{s}[c]{$\trr_s$}
  \psfrag{a}[c]{$a$}
  \psfrag{x}[c]{\underline{$a$}}
  \psfrag{b}[c]{\underline{$a\trr_s a$}}
\includegraphics[width=0.8\textwidth]{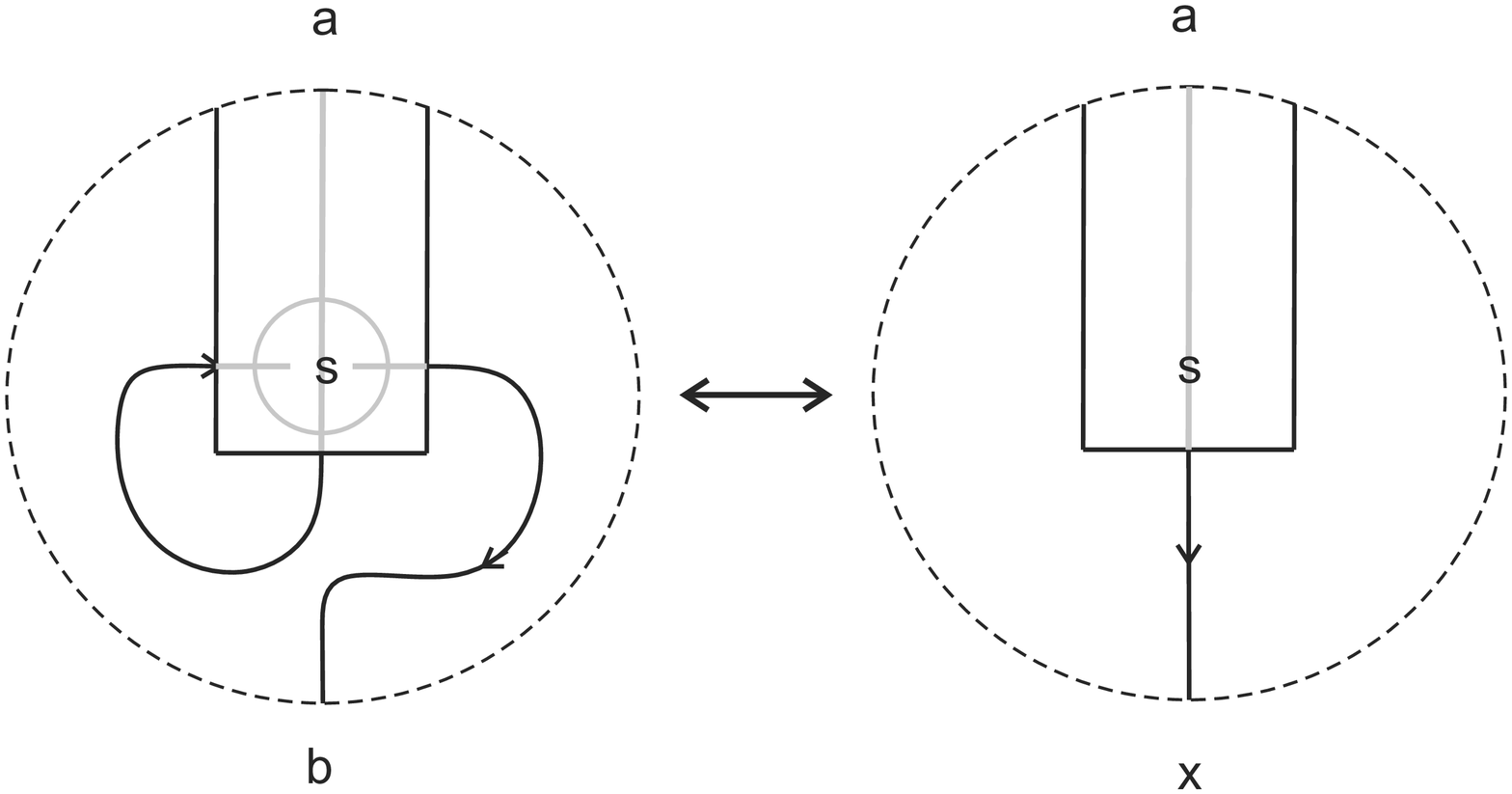}%
\caption{\label{F:r1}}
\end{subfigure}%
\quad\quad\quad
\begin{subfigure}{.5\textwidth}
  \centering
  \psfrag{c}[c]{$b$}\psfrag{a}[c]{$a$}\psfrag{x}[c]{$a$}\psfrag{d}[r]{\underline{$(a \trr_s b) \rrt_s b$}}\psfrag{b}[ld]{$a\trr_s b$}\psfrag{m}[c]{\underline{$a$}}
  \psfrag{s}[c]{$\trr_s$}
\includegraphics[width=0.8\textwidth]{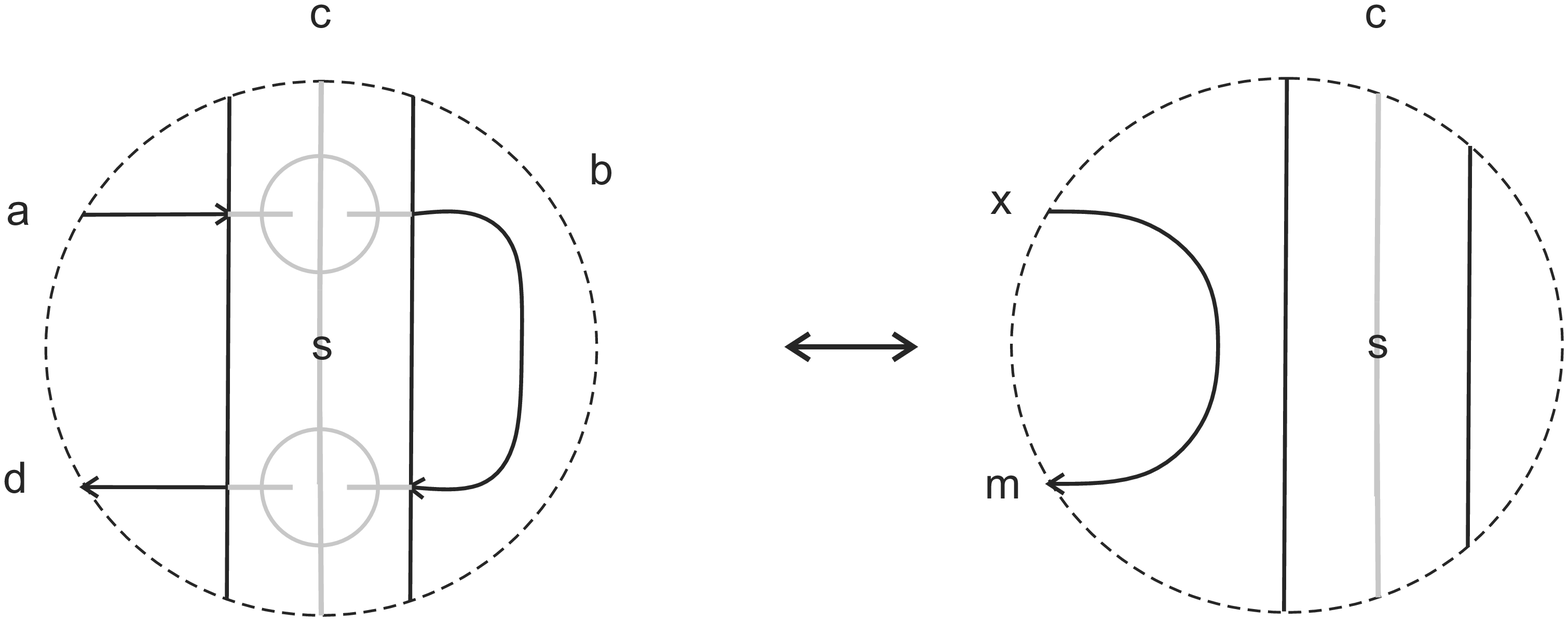}
\caption{\label{F:r2}}
\end{subfigure}%
\\[0.5cm]
\scalebox{0.99}{\begin{subfigure}{.95\textwidth}
  \centering
  \psfrag{s}[c]{$\trr_s$}
\psfrag{t}[c]{$\trr_t$}
\psfrag{a}[ld]{\underline{$(a\trr_s b)\trr_t c$}}\psfrag{b}[l]{$b\trr_t c$}
\psfrag{1}[c]{$b$}\psfrag{0}[c]{$a$}\psfrag{2}[c]{$c$}\psfrag{c}[ld]{\underline{$(a\trr_t c)\trr_s(b\trr_t c)$}}\psfrag{d}[l]{$b\trr_t c$}
\scalebox{0.9}{\includegraphics[width=0.8\textwidth]{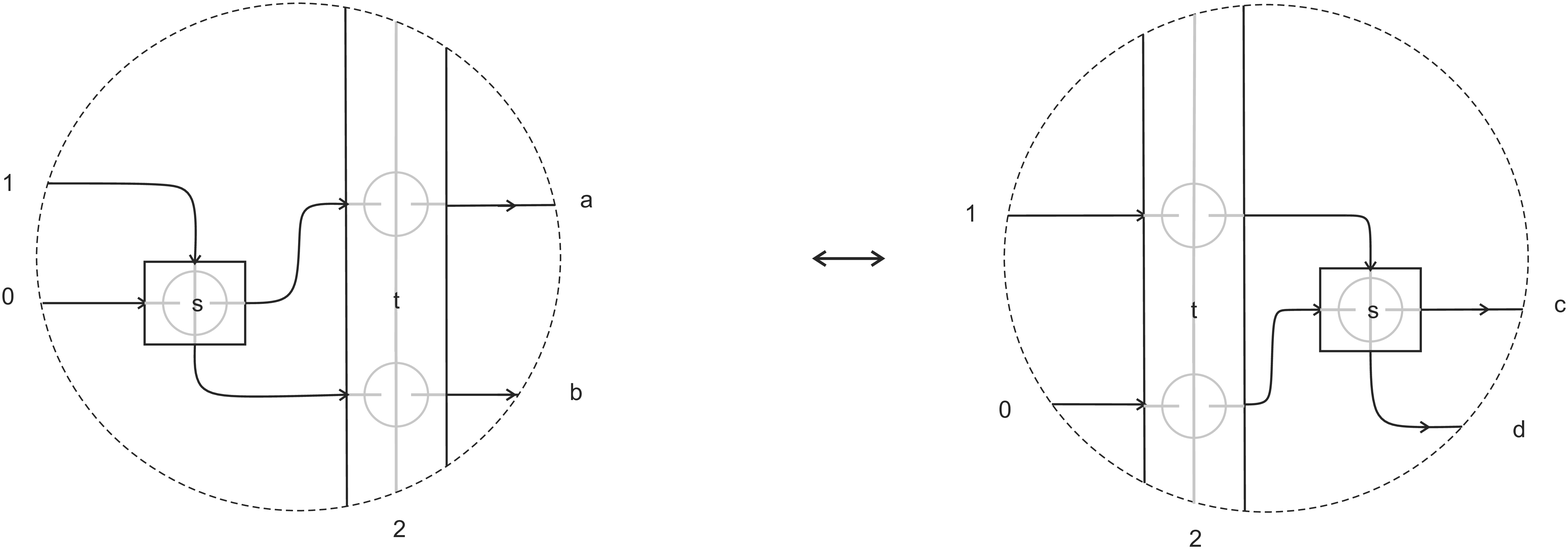}}
\caption{\label{F:r3}}
\end{subfigure}}
\caption{\label{F:ReidemeisterMoves} \small The Reidemeister Moves, which are local modifications of information fusion networks. The moves are considered for all orientations on all edges. The quandloid axioms guarantee that the underlined colours on the LHS and on the RHS are equal, and thus that these local moves are well-defined. Only a special case of R3 is illustrated above--- the general case is given by Figure~\ref{F:R3}.}
\end{figure*}

\begin{figure*}
\centering
\raisebox{0.1in}{\begin{minipage}{1.65in}
\includegraphics[width=1.65in]{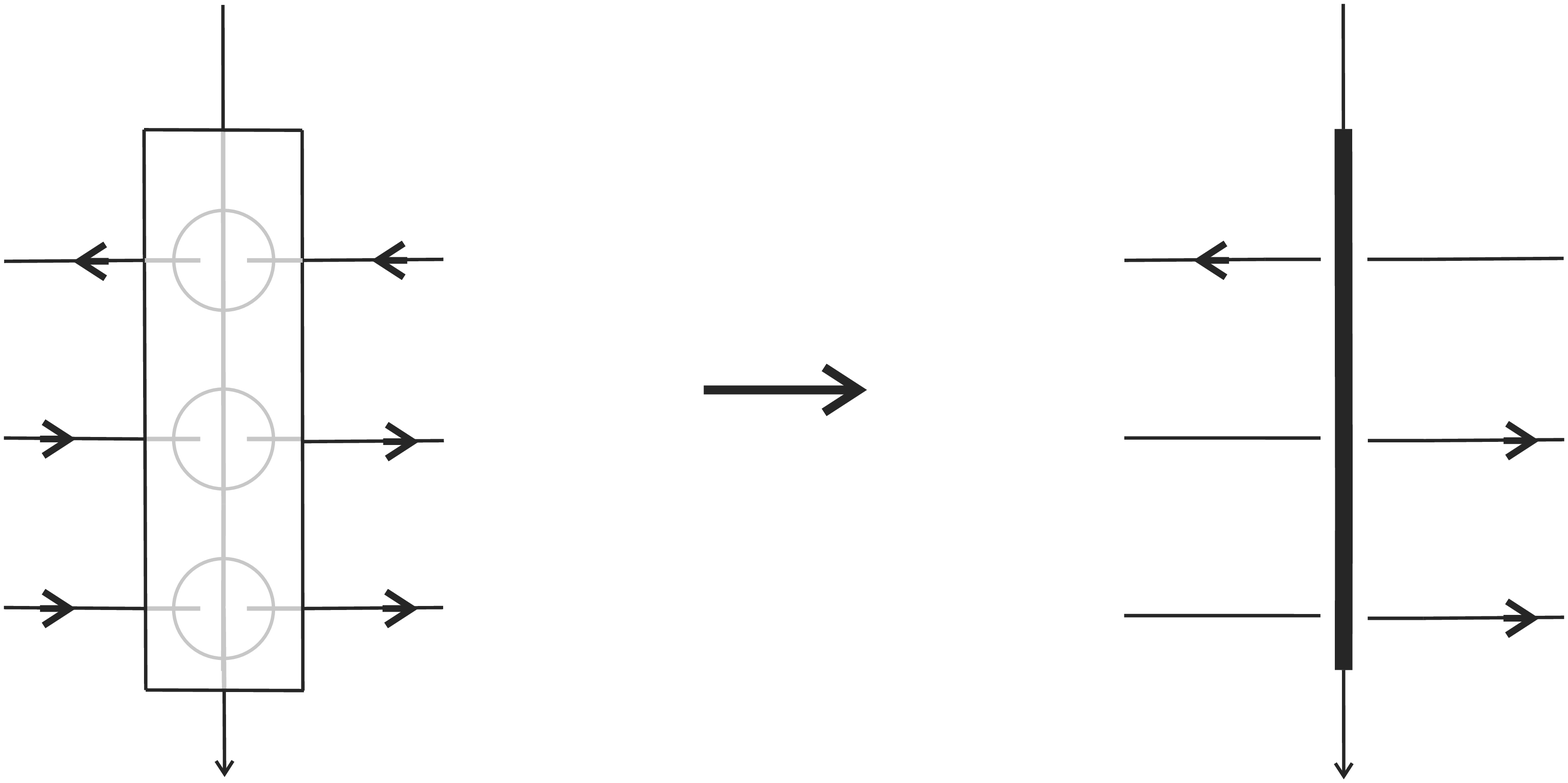}
\end{minipage}}
\qquad\qquad\quad
\begin{minipage}{2in}
\psfrag{a}[c]{\textbf{(R3)}}
\includegraphics[width=2in]{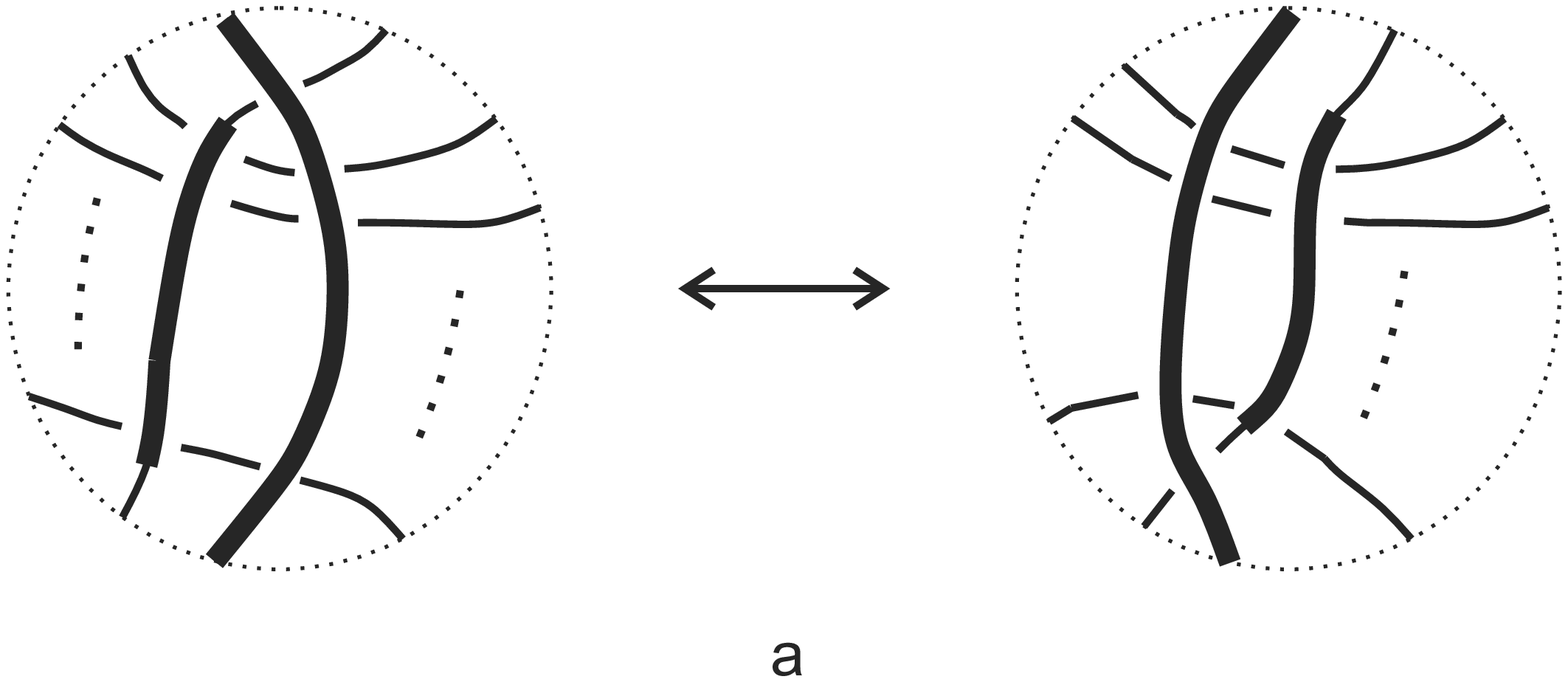}
\end{minipage}
\caption{\label{F:R3} \small For simplicity, we draw boxes by thickening the agent. We can now easily draw the R3 move.}
\end{figure*}

In this section we describe three local modifications on information fusion networks. These \emph{Reidemeister moves} are drawn in Figures~\ref{F:ReidemeisterMoves} and~\ref{F:R3}. Two information fusion networks which differ by a finite sequence of these modifications are considered to be \emph{equivalent}. For this reason, we think of these modifications as \emph{conservation laws}.




\begin{itemize}
\item \emph{Reidemeister \textrm{I}}.  This local move modifies a network by eliminating or introducing an update of a edge by itself. By coherence of information, no information is gained or lost when we perform this move.

\item \emph{Reidemeister \textrm{II}}. This local move modifies a network by updating an edge, and immediately discounting it by the same agent. By causal invertibility doing such a thing does not effect the colour of the edge, and so no information is gained or lost when we perform this move.




\item \emph{Reidemeister \textrm{III}}. This local move modifies a network by replacing updated outputs with a common agent by corresponding updated inputs updated by the correspondingly updated agent. By `no double counting' doing such a thing does not effect the colour of the edge, and so no information is gained or lost when we perform this move. 
\end{itemize}

In low dimensional topology, the \emph{Reidemeister Theorem} tells us that two diagrams of tangles are related by a finite sequence of Reidemeister moves if and only if the tangles which they represent are \emph{ambient isotopic}. Combined with a result stating that a tangle is equivalent to the set of all of its colourings, tells us that the Reidemeister moves are the unique set of local moves on tangles which preserve the information content of the tangle--- \textit{i.e.} the set of all possible tangle colourings. The appearance of Reidemeister moves signifies that the theory of information fusion networks has a low dimensional topological flavour.

We believe that the Reidemeister moves are diagrammatic representations of fundamental symmetries of information fusion (although there may also be other symmetries), and therefore that the formalism of information fusion networks subject to Reidemeister moves is a topological candidate for a suitably flexible diagrammatic language with which to discuss adaptive networks of information fusion.

\section{Topological resilience}\label{S:Resilience}


In many applications, raw data is acquired from external sources (\textit{e.g.} from sensors) over a period of time. In such cases, statistical properties of the input information are time-dependent. Suppose that we are given many such information input streams, and that our goal is to fuse some of them, maybe to discount others (perhaps in order to isolate the contribution of one set of streams over another, or perhaps for a different reason), and ultimately to generate some set of output streams. In doing so, we form a complicated information fusion network whose interaction parameters we compute to optimize some global objective function.

For example, in a network coloured by a log-linear quandloid, we may choose the \emph{Chernoff information} as our global objective \cite{Niel:13}. 
To find optimal parameters $\set{s_i}$, compute:
\begin{equation}
\label{eq:chernoff}
\min_{\{s_i\} \in (0,1)} \left[ \sum_{\{j \text{ is an output}\}} \log \int_x p_j(x) dx \right],
\end{equation}
where $p_j(x)$ denotes the unnormalized probability density in the $j$-th output.

The global goal of the network is to generate its output colours. However, the results of intermediate computations may also matter to us, for example in order to backtrack to correct an error. Reidemeister moves are local, so they change only a small part of the network, with the rest of the network remaining unchanged. At every moment the network has a `best' representation, which is the representative in our equivalence class of information fusion networks for which some chosen local objective function is maximized. The network may adaptively be configured as follows:


\begin{enumerate}
\item Optimize the network quandloid parameters $\{s_j\}$ with respect to some global objective function (e.g. \eqref{eq:chernoff}).
This defines an equivalence class of networks $[T]$.
\item Adaptively optimize the network $T^\prime \in [T]$ within its equivalence class with respect to some local objective function.
\end{enumerate}

\begin{figure*}[t]
\centering
\psfrag{s}[c]{\small $N_L$}
\psfrag{a}[c]{\small \emph{fault} $000$}
\psfrag{b}[c]{\small \emph{fault} $00X$}
\psfrag{c}[c]{\small \emph{fault} $0X0$}
\psfrag{d}[c]{\small \emph{fault} $X00$}
\psfrag{e}[c]{\small \emph{fault} $X0X$}
\psfrag{f}[c]{\small \emph{fault} $0XX$}
\psfrag{0}[c]{\small $0$}
\psfrag{1}[c]{\small $1$}
\psfrag{2}[c]{\small $2$}
\includegraphics[width=0.75\textwidth]{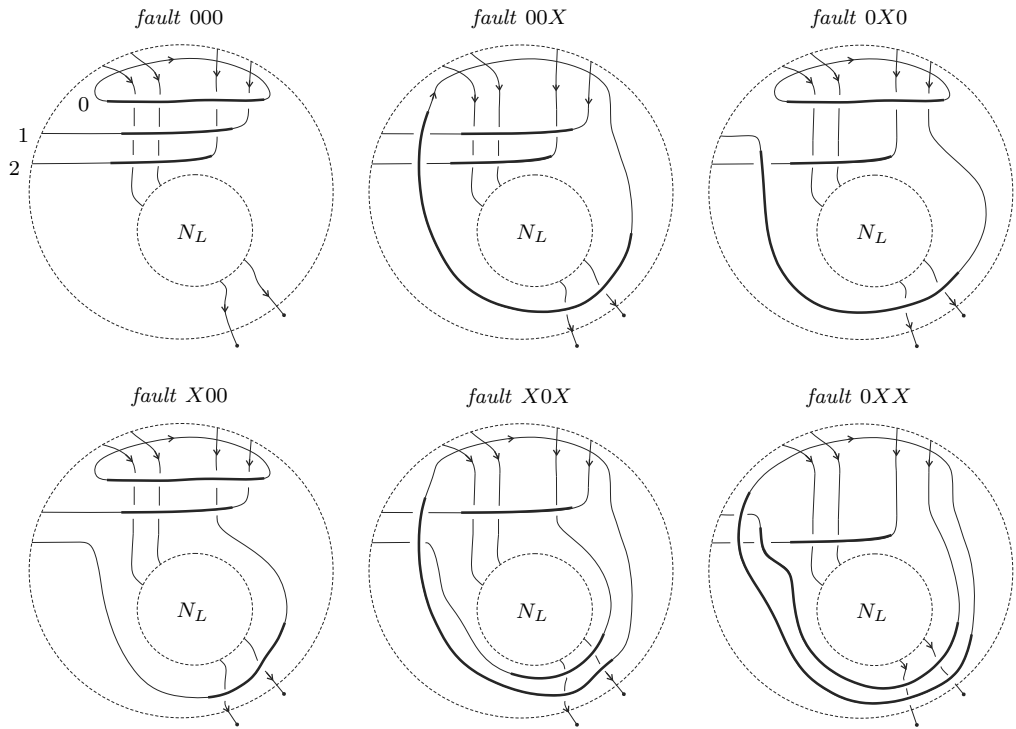}
\caption{\small Topological fault-tolerance. Equivalent networks ameliorate the effects of various occurring faults. The fault code appearing above
each network should be read as follows. The digits from left to right correspond to edges $2$,$1$ and $0$. An ``X'' in the place of the $i$th
digit signifies faulty content in its respective edge.}
\label{fig:topo}
\end{figure*}

The algorithmic aspect of designing such a topological fault-tolerance scheme to optimize a network with respect to a given local objective function might be complicated in general. To illustrate the utility of our theory, we give an example for which it is easy. Consider the network illustrated in the upper left corner of Figure~\ref{fig:topo}. This network has a set of outputs outside the bounding disk (not shown). Some set of intermediate edges lies inside the subnetwork designated by $N_L$ and represented as an empty circle.

In the course of network operation, erroneous streams of data cause one or more of the edges $0$,$1$, and $2$, to carry faulty pieces of information, \textit{e.g.} biased, inconsistent and otherwise unreliable estimates. This is detected inside the network. To inhibit the influence of this contamination on edges within $N_L$, the network performs Reidemeister moves on itself, transforming itself into an equivalent network. This is achieved by `sliding' the faulty edges all the way over $N_L$, by repeated application of the second and third Reidemeister moves. In the resulting topology, the faulty edges have no effect on $N_L$, and so local costs are improved.

\section{Conclusion}\label{S:Conclusion}

In this note, we provided an axiomatic characterization of information fusion as an operation of a \emph{quandloid}. The key properties of information fusion are no~double~counting (distributivity) and causal~invertibility. We showed that covariance intersection of estimators, fusion of observed Fisher information matrices, and fusion of Shannon information are all quandloid operations, and that they are all obtained from a single log-linear quandloid by means of homomorphisms. Inspired by the low-dimensional topological theory of tangle diagrams, we defined an \emph{information fusion network}, which comes equipped with a natural notion of equivalence. Two information fusion networks are equivalent if they differ by a finite sequence of Reidemeister moves. Our examples demonstrated that information fusion networks are fault-tolerant, in the sense that they can adaptively be optimized to minimize the effect of faulty input information on the network.

\appendix

\subsection{Proof of Proposition~\ref{thm:ci}}

The covariance intersection update rules are obtained via a homomorphism from a particular log-linear quandloid.

\begin{proof}
Let $p(X), p(Y) \in Q$ be unnormalized Gaussian densities parameterized by their mean and covariance, $(\hat{X}, C_X)$ and $(\hat{Y}, C_Y)$, respectively.
An update operation reads:
\begin{multline}
\label{eq:zz}
p(X) \trr_s p(Y)=  p(Z) \\
= \exp\bigg(-\frac{1}{2} \bigg[ (1-s) (Z - \hat{X})^T C_X^{-1} (Z - \hat{X}) \\
+ s (Z - \hat{Y})^T C_Y^{-1} (Z - \hat{Y})\bigg]\bigg)
\end{multline}

We verify that this expression equals:
\begin{equation}
\exp\left(-\frac{1}{2} (Z - \hat{Z})^T C_Z^{-1} (Z - \hat{Z}) \right),
\end{equation}
for $\hat{Z}$ the form which appears in \eqref{E:21} and for $C_Z$ of the form which appears in~\eqref{E:22}.

Given $p(X) \in Q$, we compute that $C_X^{-1} = -\frac{\partial^2 \log p(x)}{\partial x \partial x^T}$, which is called the \emph{precision matrix} of $p(X)$. In particular,
\begin{equation}
C_Z^{-1} = - \frac{\partial^2 \log p(z)}{\partial z \partial z^T} = (1-s) C_x^{-1} + s C_Y^{-1} = C_x^{-1} \bar{\trr}_s C_Y^{-1}
\end{equation}
which follows from \eqref{eq:zz}.


For any $p(X) \in Q$ the mode $\hat{X}$ satisfies:
\begin{equation}
 \left.\frac{\partial \log p(x)}{\partial x} \right|_{x=\hat{X}} = 0
\end{equation}
The mode $\hat{Z}$ of $p(Z)$ is thus obtained from~\eqref{eq:zz} by
\begin{equation}
2 \left.\frac{\partial \log p(z)}{\partial z} \right|_{z=\hat{Z}} = (1-s) C_x^{-1} (\hat{Z} - \hat{X}) + s C_Y^{-1} (\hat{Z} - \hat{Y}) = 0
\end{equation}
finally yielding
\begin{equation}
C_z^{-1} \hat{Z} = (1-s) C_x^{-1} \hat{X} + s C_Y^{-1} \hat{Y} = \left(C_x^{-1} \hat{X}\right) \bar{\trr}_s \left(C_Y^{-1} \hat{Y}\right).
\end{equation}
\end{proof}


\end{document}